\documentclass[prl,twocolumn,showpacs,amsmath,amssymb,superscriptaddress,preprintnumbers]{revtex4}
\usepackage{graphicx}
\begin{document}

\preprint{JLAB-THY-09-1034, ANL-PHY-12435-TH-2009}

\newcommand{\mev}{\,{\rm MeV}}
\newcommand{\ket}[1]{{ | #1 \rangle }}
\newcommand{\bra}[1]{{ \langle #1 | }}
\newcommand{\s}{{\sigma}}
\newcommand{\beq}{\begin{eqnarray}}
\newcommand{\eeq}{\end{eqnarray}}
\newcommand{\scinot}[2]{#1 \times 10^{#2}}
\newcommand{\scinott}[2]{$\scinot{#1}{#2}$}

\renewcommand{\thefootnote}{\fnsymbol{footnote}}

\author{Joel Giedt\footnote{{\tt giedtj@rpi.edu}}}
\affiliation{Department of Physics, Applied Physics, and Astronomy,
    Rensselaer Polytechnic Institute,
    110 Eighth Street, Troy, New York 12180-3590 USA}
\author{Anthony W.~Thomas\footnote{{\tt awthomas@jlab.org}}}
\affiliation{Jefferson Lab, 12000 Jefferson Avenue, Newport News, Virginia 23606, USA \\
College of William and Mary, Williamsburg, Viginia 23187, USA}
\author{Ross D.~Young\footnote{{\tt ryoung@anl.gov}}}
\affiliation{Physics Division, Argonne National Laboratory, Argonne, Illinois 60439, USA}

\title{Dark matter, the CMSSM and lattice QCD}
\begin{abstract}
  Recent lattice measurements have given accurate estimates of the
  light and strange quark condensates in the proton.  We use these new
  results to significantly improve the dark matter predictions in a
  set of benchmark models that represent different scenarios in the
  constrained minimal supersymmetric standard model (CMSSM). Because
  the predicted cross sections are at least an order of
  magnitude smaller than previously suggested, our results have
  significant consequences for dark matter searches.
\end{abstract}

\maketitle

\renewcommand{\thefootnote}{\arabic{footnote}}

The astronomical evidence for dark matter~\cite{Jarosik:2006ib,Tegmark:2003ud} 
presents modern physics with 
some of its greatest challenges. We need to find ways to detect it directly 
and a number of very sophisticated searches are 
underway~\cite{Ahmed:2008eu,Angle:2008we,Aalseth:2008rx}, 
with at times confusing results~\cite{Bernabei:2008yi,Savage:2009mk}. 
To guide those searches we need theoretical models for what the dark matter
might be and what cross section one might expect it to have for scattering from 
hadronic matter. The constrained minimal supersymmetric extensions
of the Standard Model (CMSSM---see \cite{deBoer:1994dg} for a review)
 have the advantage that they are consistent 
with modern nuclear and particle physics while incorporating the compelling 
concepts of supersymmetry, coupling unification,
and viable cold dark matter~\cite{Jungman:1995df}. 
These models have over time been tuned to 
ensure that they are consistent with the latest constraints on relic 
abundance~\cite{Battaglia:2001zp,Battaglia:2003ab,Ellis:2008hf,Ellis:2009ai}, with 
the favoured dark matter candidate being a neutralino with mass of order 100 GeV 
and a density of a few per litre. 

Extensive studies of the spin-independent interaction of neutralinos with 
hadronic matter have established that the cross section is 
determined by their    
scalar coupling to the light and strange quark sigma commutators 
($(m_u + m_d) \langle \bar{u} u + \bar{d} d \rangle / 2$ and 
$ m_s \langle \bar{s} s \rangle $); 
see~\cite{Ellis:2008hf,Ellis:2009ai} and references therein. 
Given that the favoured values of these quantities have been typically 
50 MeV~\cite{Gasser:1990ce} and 300 MeV~\cite{Nelson:1987dg}, 
respectively, the dark matter cross section has been 
dominated by the strange quark content of the proton. Realizing the 
importance of these quantities Ellis {\it et al.} recently made a plea for  
more accurate experimental data for them~\cite{Ellis:2008hf}. 
While this seems unlikely in the short term, at least, 
the answer to their plea has recently been provided 
from an unexpected source, through the study of octet baryon masses 
as a function of quark mass. Indeed, a sophisticated chiral analysis of     
recent lattice measurements has yielded surprisingly accurate 
estimates of these sigma commutators~\cite{Young:2009zb}.
Subsequent lattice computations have found 
consistent results \cite{Toussaint:2009pz}.

In this  Letter, we use the precise new values of the
light and strange quark 
sigma commutators obtained in Refs.~\cite{Young:2009zb,Toussaint:2009pz} to
update the cross sections 
predicted within the benchmark CMSSM models that have been studied for
several years by Ellis {\it et al.}~\cite{Ellis:2008hf}
and their collaborators~\cite{Battaglia:2001zp,Battaglia:2003ab}.
We show that the sizeable reduction in the strange sigma term
from the values previously favoured leads to 
a rather dramatic reduction in the expected neutralino cross sections, with
important implications for the interpretation of experiments to search for 
dark matter.

\paragraph{Sigma terms from lattice QCD.}
The sigma terms of the nucleon, for each quark flavour, are given by
the scalar form factors evaluated at $t=0$, denoted by
\begin{equation}
\sigma_q = m_q \langle N | \bar{q} q | N \rangle .
\end{equation}
The nucleon scalar form factor is difficult to directly probe in
experiment. The light-quark sigma term, $\sigma_\ell=(m_u+m_d)\langle
\bar{u}u+\bar{d}d\rangle/2$, has most reliably been accessed by
invoking a chiral low-energy relation between $\pi$--$N$ scattering
and the scalar form factor
\cite{Cheng:1970mx,Brown:1971pn,Gasser:1990ce},
\begin{equation}
\Sigma_{\pi N} \equiv \sigma_\ell 
            = \Sigma_{\pi N}^{\rm CD} - \Delta_R - \Delta_\sigma \,.
\end{equation}
The remainder term, $\Delta_R$, describes a
correction to the low-energy theorem and is estimated to be less than
$2\mev$ \cite{Brown:1971pn,Bernard:1996nu}.  The shift in the scalar
form factor can be inferred from a dispersion analysis and found
to be rather large \cite{Gasser:1990ap},
$\Delta_\sigma \equiv \sigma_\ell(2m_\pi^2) - \sigma_\ell = 15.4 \pm 0.4 \mev$.
$\Sigma_{\pi N}^{\rm CD}$ is the Born-subtracted, isoscalar $\pi N$
scattering amplitude evaluated at the (unphysical) Cheng-Dashen
point. An early experimental extraction \cite{Koch:1982pu} gave 
$\Sigma_{\pi N}^{\rm CD}= 64 \pm 8\mev$, to be compared with more a
recent determination $\Sigma_{\pi N}^{\rm CD}=79\pm7\mev$
\cite{Pavan:2001wz}. These two values lead to light-quark sigma terms
of
\begin{equation}
\sigma_\ell = 45 \pm 8\mev \quad \text{and} \quad 64 \pm 7\mev,
\end{equation}
respectively. These are to be compared with the recent lattice QCD determination 
$\sigma_\ell=47\pm9\mev$ \cite{Young:2009zb}. While this lattice
analysis tends to favour the lower value, it is not inconsistent with
the higher extraction. We use the lattice determination in the
current analysis.

Extracting the strangeness sigma term is significantly more
challenging, because the same prescription would lead to both a poorly
converged low-energy relation and a large extrapolation of $K$--$N$
scattering to the Cheng-Dashen point --- see Reya
\cite{Reya:1974gk}. A much more practical approach has been to resolve
the patterns of SU(3) breaking among the baryon octet
\cite{Gasser:1982ap,Nelson:1987dg,Borasoy:1996bx}. In essence, the
baryon masses give guidance with respect to the symmetry breaking
component
\begin{equation}
\sigma_0 = m_\ell \langle N | \bar{u}u + \bar{d}d - 2\bar{s}s | N \rangle\,,
\end{equation}
with $m_\ell=(m_u+m_d)/2$. The commonly reported value
$\sigma_0=36\pm7\mev$ is based on the phenomenological analysis of
corrections to the linear mass-splittings relations
\cite{Gasser:1982ap}, with further uncertainties from yet higher order
quantified by \cite{Borasoy:1996bx}.  The difference between the
estimated $\sigma_0$ and the extracted $\sigma_\ell$ then gives a
best estimate for the strange-quark sigma term, as related by
\begin{equation}
\frac{\s_s}{\s_\ell} = \frac{m_s}{2m_\ell} \left(
\frac{\Sigma_{\pi N}- \s_0}{\Sigma_{\pi N}} \right).
\label{jeq}
\end{equation}

In contrast, lattice QCD allows one to directly study the quark mass
dependence of the baryon masses from first principles. New lattice
results, based on an SU(3) baryon mass analysis \cite{Young:2009zb}
and a novel application of the Feynman-Hellman theorem at the
correlator level \cite{Toussaint:2009pz}, find $\sigma_s=31\pm15\mev$
and $59\pm10\mev$, respectively. An important feature of these
calculations is the demonstrated internal consistency, where both
studies have successfully shown the ability to predict the
mass-splittings between simulations with different strange quark mass
parameters. We argue that these results show a vast improvement over the
earlier phenomenological analyses. This, in turn, implies that it is 
appropriate to update the earlier estimates of
dark matter cross sections. 
In the present analysis, we assume that the reported uncertainties 
in the lattice QCD determinations of the strange quark sigma term are 
independent and use a na\"ive average of the two results,
$\sigma_s=50\pm8\mev$. We report the results of 
our analysis at the 95\% confidence
level to ensure that the conclusions are minimally sensitive to our
input.

\paragraph{Constrained MSSM.}
The constrained minimal supersymmetric standard model (CMSSM) is
inspired by supergravity mediated scenarios of spontaneous
supersymmetry breaking, in the context of supersymmetric grand unified
theories \cite{deBoer:1994dg}.  Then one has a 
universal soft scalar mass $m_0$, a
universal gaugino mass $m_{1/2}$, a universal trilinear scalar
coupling $A_0$ (which is set to zero in all the models considered
here), and a higgsino mass $\mu$.  In practice one specifies the ratio
of vacuum expectation values $\tan\beta=v_u/v_d$ of the Higgs fields
$H_u$ and $H_d$ at the low energy scale ($m_Z$) and determines $\mu$
from this constraint; the sign of $\mu$ is also a parameter, and
unification of gauge couplings is imposed.  One advantage of the CMSSM
class of models is its simplicity, allowing for detailed scans over
parameter space.  The benchmark models ``A-M'' of
\cite{Battaglia:2001zp,Battaglia:2003ab} were selected to be
representative of the regions of parameter space that yield neutralino
dark matter with the correct abundance, consistent with WMAP
constraints.

Because of the incredible accuracy of the WMAP results
for relic cold dark matter abundance, a fine tuning
of high scale parameters is typically required
in order to have the lightest supersymmetric particle (LSP) fit this constraint.  However,
the dark matter cross section itself only changes
by a few percent when this tuning is done.  For instance,
the tuning that we made (described below) in Model L, which is a 5\%
change to $m_0$, only results in a 2\% change
in the spin-independent cross section.  Thus
the direct detection predictions are insensitive
to the changes that may have to be made to bring
the relic abundance into line with any future updates.
Thus the cross section results presented here are
robust in that sense.

\paragraph{Cross sections for benchmark models.}
We have computed the spin-independent cross section
for neutralino dark matter for Benchmark Models A-M of \cite{Battaglia:2001zp,
Battaglia:2003ab}.  Three of the benchmark models 
were also studied in \cite{Ellis:2008hf}, where
the reader may find the cross section formulae that
we likewise use.  While Ellis et al.~use a private
renormalization group evolution code,
we have used the publicly available Softsusy \cite{Allanach:2001kg}
to compute the running parameters between the grand
unified scale and the electroweak scale.  (We did, however,
compare our Softsusy results with those obtained using
the code of Ellis et al.~and found good agreement.)
Minor modifications were necessary in two benchmark models,
because of the slight differences between the renormalization group codes.
These small shifts in high scale inputs were made
in order to avoid a stau ($\tilde \tau$) 
LSP for models that are on the
edge of the $\tilde \tau$ exclusion region of CMSSM (i.e.,
dark matter should be neutral).
In Model J we changed $m_0$ from 285 GeV to 290 GeV.
In Model L we changed $m_0$ from 300 to 315 GeV (305 and 310 were
insufficient).  As discussed above, these changes have
an insignificant effect on the cross section values.

Our results for the spin independent cross section with the proton are
shown in Figs.~1-3.  In all cases we see that there is only a mild
dependence on $\Sigma_{\pi N}$.  Furthermore, most models have rather
small cross sections, $\s_{SI} < 10^{-9}$ pb.  We do not show Model L
because it is essentially degenerate with Model I.  Nor do we show
Model D because it has such small cross sections ($\s_{SI} < 10^{-14}$
pb) that it is unobservable in the forseeable future.  We also
computed the cross sections for the neutron and found them to be
essentially degenerate with those of the proton.  By way of contrast with
previous estimates, in Fig.~4 we show $\s_{SI}$ for Model C using
$\s_\ell$ and $\s_s$ extracted from lattice QCD with the
determination of $\s_s$ inferred from the phenomenological estimate
for $\s_0 \, = \, 36$ MeV, using Eq.~(\ref{jeq}) above.  
Not only is the cross section
calculated here, on the basis of the latest lattice QCD results, far 
more weakly dependent on $\Sigma_{\pi N}$, but it is
also typically much smaller.

\begin{figure}
\begin{center}
\includegraphics[width=3in,height=2.5in,angle=90]{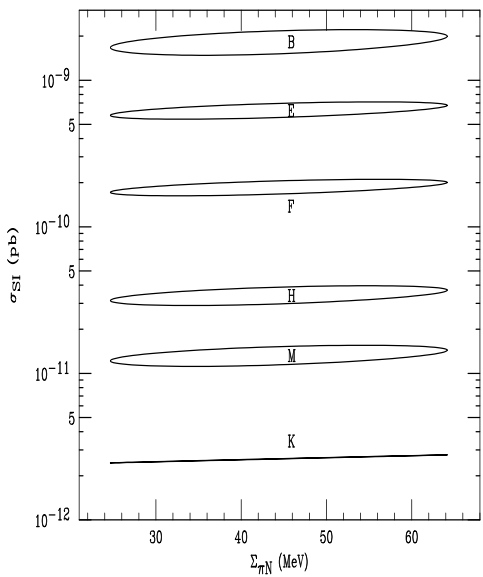}
\caption{Proton, benchmark models (labeled by letters A, B,...), 95\% CL using lattice inputs
for sigma commutators.  Remaining models are
shown in other figures below, since they would overlap with these.
The neutron predictions are not shown because they are
basically degenerate with the proton cross sections on the logarithmic
scale that is shown.}
\end{center}
\end{figure}

\begin{figure}
\begin{center}
\includegraphics[width=3in,height=2.5in,angle=90]{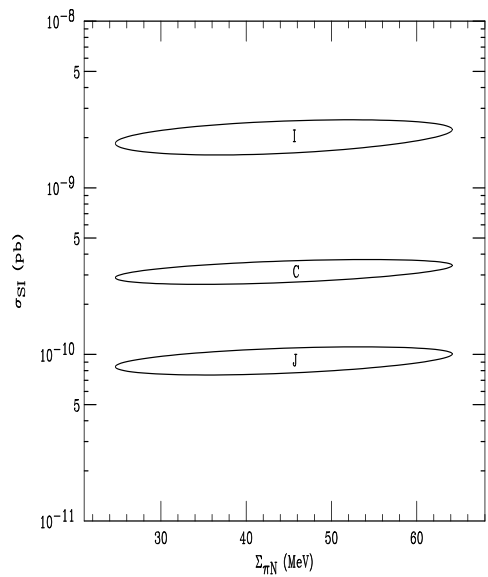}
\caption{Cross section estimates (95\% CL) for benchmark models C, I, J.}
\end{center}
\end{figure}

\begin{figure}
\begin{center}
\includegraphics[width=3in,height=2.5in,angle=90]{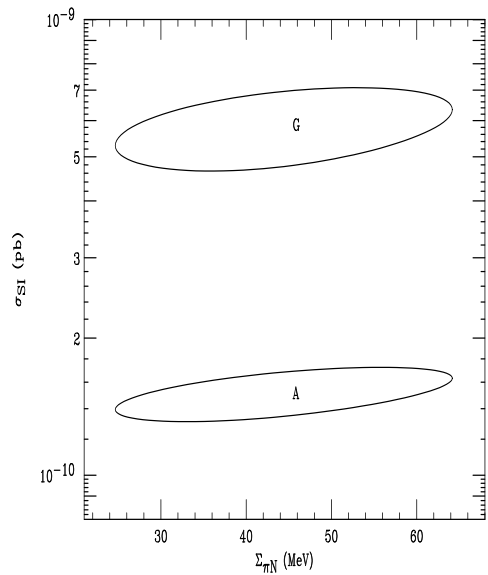}
\caption{Cross section estimates (95\% CL) for benchmark models A, G.}
\end{center}
\end{figure}

\begin{figure}
\begin{center}
\includegraphics[width=3in,height=2.5in,angle=90]{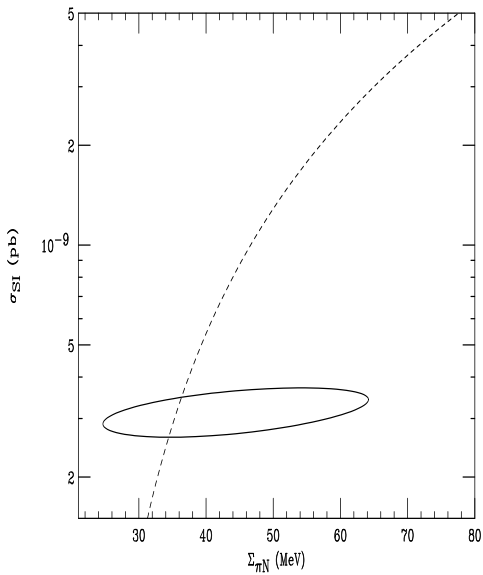}
\caption{A comparison of our results (solid ellipse) for Model C, versus
the traditional approach (dashed line) which relates the strange quark
sigma commutator to the light quark one through Eq.~(\ref{jeq})
with $\sigma_0 = 36$ MeV.
\label{figD}
}
\end{center}
\end{figure}

\paragraph{Details of quark flavor contributions.}
It is interesting to investigate the breakdown of how each quark
flavor contributes to the overall cross section.  First,
we note that for the proton $\s_{SI} \propto |f_p|^2$,
where
\beq
&& m_p f_p = \sum_{q=u,d,s} \frac{\alpha_{3q}}{m_q} f_{Tq}^p
+ \frac{2}{27} f_{TG}^p \sum_{q=c,b,t} \frac{\alpha_{3q}}{m_q},
\nonumber \\
&& f_{TG}^p = 1 - \sum_{q=u,d,s} f_{Tq}^p
\label{ameq}
\eeq
The coefficients $\alpha_{3q}$ are determined by squark
exchange and mixing coefficients of the LSP neutralino.
The dimensionless sigma commutators $f_{Tq}^p = \s_q/m_p$
are taken from the lattice results, as explained earlier, 
whereas $f_{TG}^p$ addresses
the heavy flavor sigma commutators through SVZ relations,
as discussed in the review \cite{Jungman:1995df}.  We also find it interesting
to compute the relative contribution of each quark flavor,
defining $f_p = \sum_{q} f_q^p$.  For instance, the
bottom quark contribution $f_b^p=
(2/27) m_p f_{TG}^p \alpha_{3b}/m_b$ turns out to
be rather large.  Its increased importance in our
computations can be traced to the
fact that the lattice results give a smaller result for
$f_{Ts}^p$, the dimensionless strange sigma commutator.
This enhances $f_{TG}^p$, as can be seen from Eq.~(\ref{ameq}).
Results are given in Table \ref{tab1} for the point with
maximim $\s_{SI}$, which occurs in Model L.

\begin{table}
\begin{center}
\begin{ruledtabular}
\begin{tabular}{c|c|c|c|c}
Model & $q$ & $\alpha_{3q}/m_q$ & $f_{Tq}^p$ or $f_{TG}^p$ & $f_q^p/f_p$ \\ \hline
L (max $\s_{SI}$)             & u & \scinott{-1.019}{-09} & 0.0280 & 0.0105 \\
$\s_{SI} =$                   & d & \scinott{-1.302}{-08} & ''     & 0.1342 \\
$\quad 2.8 \times 10^{-9}$ pb & c & \scinott{-1.031}{-09} & 0.8751 & 0.0261 \\
                              & s & \scinott{-1.522}{-08} & 0.0689 & 0.3633 \\
                              & t & \scinott{-1.936}{-09} & 0.8751 & 0.0462 \\
                              & b & \scinott{-1.670}{-08} &  ''    & 0.3984 \\
\hline
\end{tabular}
\end{ruledtabular}
\caption{Example breakdown of quark flavor contributions to $\s_{SI}$.
This is for Model L at the maximum value of its cross section,
within our 95\% CL region.
\label{tab1}}
\end{center}
\end{table}

\paragraph{Conclusions.}
In this Letter we have shown that the recent lattice results \cite{Young:2009zb,Toussaint:2009pz}
for $\s_\ell$ and $\s_s$
have a dramatic effect on predictions for
direct detection of neutralino cold dark matter.    Furthermore, the theoretical
uncertainty is considerably smaller than in the traditional approach,
as was highlighted in Fig.~4.  Unfortunately, the most
important effect of the improved estimates is the reduction
of the strange quark sigma commutator, which leads to
rather small dark matter cross sections. 
On the other hand, even within the 95\% confidence intervals, our
estimates come with small errors; hence any observation of dark matter
would be highly selective amongst the benchmark models that we have
considered.  In future work we hope to report on dark matter cross
section predictions in models other than the CMSSM, such as the
nonuniversal Higgs mass (NUHM) models, which are known
\cite{Ellis:2009ai} to allow for cross sections which are larger than
in the CMSSM.

\paragraph{Acknowledgements.}
We thank Keith Olive for helpful discussions and for allowing us to
use a version of his collaborations' RGE code in order to perform
comparisons to the Softsusy output.  JG was supported by Rensselaer
faculty development funds.  This work was supported by the the
U.S. Department of Energy contracts DE-AC05-06OR23177, under which
Jefferson Science Associates, LLC operates Jefferson Laboratory and
DE-AC02-06CH11357, under which UChicago Argonne, LLC, operates Argonne
National Laboratory.

\end{document}